\documentclass[a4paper,twoside]{article}
\newcommand{\affil}[1]{$^{\rm #1}$}
\usepackage[authoryear]{natbib}
\bibpunct{(}{)}{;}{a}{}{,}
\usepackage{graphicx}
\date{} %Please leave the date blank
\title{\large\bf\flushleft The Distance to NGC 5128 (Centaurus A) }
\author{\parbox{\textwidth}{\flushleft
\vspace{-0.5cm}
{\it Gretchen L. H. Harris\affil{A,D}, Marina Rejkuba\affil{B}, and William 
E. Harris\affil{C}}\\
\vspace{0.4cm}
{\small \affil{A}\,Department of Physics and Astronomy, University of Waterloo, Waterloo ON N2L 3G1, Canada}\\
{\small \affil{B}\,European Southern Observatory, Karl-Schwarzschild-Strasse 2, D-85748 Garching, Germany}\\
{\small \affil{C}\,Department of Physics and Astronomy, McMaster University,
Hamilton ON, L8P 2X7, Canada}\\
{\small \affil{D}\,Email: glharris@astro.uwaterloo.ca}}}
\begin{document}
\twocolumn[
\begin{changemargin}{.8cm}{.5cm}
\begin{minipage}{.9\textwidth}
\vspace{-1cm}
\maketitle
%
%
%%%%%%%%%%%%%     ABSTRACT    %%%%%%%%%%%%%
%Abstract of no more than 200 words here.
\small{\bf Abstract:}

In this paper we review the various high precision methods 
that are now available to determine 
the distance to NGC 5128.  These methods include: 
Cepheids, TRGB (tip of the red giant branch), 
PNLF (planetary nebula luminosity function), 
SBF (surface brightness fluctuations)
and Long Period Variable (LPV) Mira stars.  
From an evaluation of these methods 
and their uncertainties, we derive a best-estimate 
distance of $3.8 {\pm}0.1$ Mpc to NGC 5128 and find that
this mean is now well supported by the current data.  
We also discuss the role of NGC 5128 more generally 
for the extragalactic distance
scale as a testbed for the most direct possible comparison
among these key methods.

%%%%%%%%%%%%%     KEYWORDS    %%%%%%%%%%%%%
\medskip{\bf Keywords:} Galaxies: distances and redshifts --- galaxies:
stellar content --- galaxies: individual (NGC 5128)
% Please write all keywords in lower case. PASA uses the
% standard list of subject headings adopted by The Astrophysical Journal
% and available from http://www.journals.uchicago.edu/ApJ/keywords_text.html.
% Keywords are separated by em-dashes, i.e. ---

%%%%%%%%DO NOT EDIT%%%%%%%%%%%%
\medskip
\medskip
\end{minipage}
\end{changemargin}
]
\small
%%%%%%%%EDIT FROM HERE%%%%%%%%%%%%

\section{Introduction}
%Please see the PASA Style Guide for help with correct layout for your manuscript.
%Examples of tables and figures are given below.

An issue arising frequently in the literature on NGC 5128
is a lack of consensus
%%the {\it Many Faces of Centaurus A} conference was a lack of consensus
about the best distance to adopt for this important nearby elliptical galaxy. 
As with any other galaxy, the assumed distance directly affects the
astrophysics of all its components through the 
calculation of every intrinsic scale length and every luminosity,
at every wavelength.  Some researchers use 
%%Some participants at the conference used   
a distance of 3.9 Mpc from a weighted average of several methods which measure
 the properties of old resolved stars \citep{rej04}. 
Others quote a value of d = 3.4 Mpc which is based only on 
Cepheids \citep{fer07}. 
This uncomfortable uncertainty of almost 20\% is much larger than 
one would like for a galaxy we all agree is a keystone object  
regardless of the wavelength region in which we work.  
Clearly the need exists for some kind of ``best estimate'' distance.

Fortunately there are now several distance indicators available to us,
and the time is right to construct a useful mean distance measurement.
None of these methods on its own can be considered by definition 
to give the {\it correct} value, because each is limited by its own
set of systematic (and random) uncertainties.
The strategic advantage of combining several methods is that it
will allow us to use many different stellar components of the galaxy,
each with its distinctive merits and each of which is at least
partly independent of the others.

\section{Evaluation of Five Methods}

At present we have five ``standard candles'' for NGC 5128 that
refer directly to properties of its stars in various ways.
These methods include:  Cepheids; the magnitude of the tip of the red 
giant branch of the oldest halo stars (TRGB); the planetary 
nebula luminosity function (PNLF); surface brightness 
fluctuations (SBF); and long period variables 
(LPV/Miras).  Below we briefly describe these methods 
to give a sense of some of the issues involved and summarize
the recent history of the subject. 

A brief statement of the philosophical approach we will take
in the discussion is appropriate at this point.
In principle, we would expect each of these methods to give us the same 
distance to NGC 5128 within their measurement uncertainties.
If those methods agree, then our confidence 
in the result is high.  But, {\sl that confidence will be stronger if the methods
are independent of each other}, i.e. if they have 
different astrophysical underpinnings
 and different means of calibrating their zero points.  For instance, if  
we find the same distance based on two methods with the same calibration
 base, then what we mostly know is that both methods have been applied 
self-consistently and appropriately.  On the other hand, good
agreement between two {\sl independent} methods tells us more than  
that those methods are reliable.  An important side benefit is
that it also increases our confidence
in cases where only one of them can be used in another galaxy. 
Conversely, disagreement tells us that the astrophysical or calibration
foundation of one or the other, or both, needs reexamination.

In several previous papers on the extragalactic distance scale
there was a tendency to normalize all calibrations to 
the prominent Cepheid method.  But such an approach
loses sight of the different strengths of the other methods and
places reliance of the local distance scale too strongly on one
method, creating a kind of distance-scale monoculture.  Our view is that 
the use of multiple methods that are as independent as possible is 
the best way to proceed.   

NGC 5128 provides a testbed for five different methods, four of which
rely on observations of resolved stars.  
In future it may be possible to add other stellar standard candles 
to the list, including RR Lyrae variables, RGB clump magnitude, 
blue supergiants, Population II Cepheids, novae, or eclipsing binaries.
But, at present, no data of sufficient quality for distance determination
exist for such objects within NGC 5128.

\subsection{Cepheids}

Distances based on Cepheids use the well-known relation between 
pulsation period and luminosity \citep[Leavitt law;][]{leavitt12}
to infer the luminosity of Population I Cepheid variable stars. 
One of the major results from the {\it Hubble Space Telescope} (HST), 
and certainly among the most well known is the determination of the
Hubble constant in the $H_0$ Key Project \citep{kennicutt+95,freedman+01}.
As described by \citet{fer00} this project depended heavily on 
Cepheids, and was designed to establish Cepheids as the primary
standard candle. The great success of the Key Project and the
attention paid to it
resulted also in wide acceptance of Cepheids as reliable
distance indicators.

However in the last decade, the number of papers discussing the influence
of chemical composition, and on-going debate on dependence or independence
of the Cepheid period-luminosity (PL) 
relation on metallicity (with discrepant results even about
the sign of the dependence on iron abundance) continues to be high, 
culminating in a recent claim by
\citet{romaniello+08} that the Cepheid PL relation is
not universal.  In addition there are other 
discussions and criticisms (some rejected) 
of biases due to blending in distant crowded 
extragalactic fields, the longest observable period, dust extinction or 
cut-off due to limiting absolute magnitude of the Cepheid sample
\citep[e.g.\ ][]{bresolin+05, paturel05, tammann+08}. 
The successful comparison of Cepheids with other standard candles 
\citep[e.g.\ ][]{rizzi+07,mould+sakai09}
and numerous Cepheid distance scale calibrations
with a very wide range of methods, such as theoretical 
non-linear pulsation models, 
parallaxes (from both Hipparcos and from HST), the Baade-Wesselink method, 
the hydrogen maser distance to NGC 4258, and main sequence fitting, 
certainly confirms that Cepheids are a good distance indicator. 
But the above mentioned 
limitations and the possibility of a non-universal PL relation 
should warn the reader that Cepheids are
not flawless standard candles and -- as for any other standard
candle -- the quoted distances have to be taken 
into account with their
error-bars included.

As already mentioned, the primary uncertainties in Cepheid distances
are reddening and dependence of the PL relation on chemical abundance. 
Both factors are important in this case. The NGC~5128 Cepheids 
are found in and near the central dust lane because Cepheids  
are relatively young stars and would not be present 
in the old (gas poor) stellar halo of an E galaxy.
Their individual reddenings are large, 
ranging from $E(V-I)\simeq$ 0.4 to 0.8 and, 
combined with somewhat uncertain values for the ratio of total 
to selective absorption, can lead to larger uncertainties in 
the dereddened absolute magnitudes. In addition, NGC~5128 
is the first elliptical galaxy in which Cepheids have been 
identified, making the environment somewhat unusual. 
\citet{fer07} quote a Cepheid distance modulus 
of $(m-M)_0 = 27.67 \pm 0.12/0.16$; the first (smaller) 
 uncertainty is random and the second is systematic.  
For straightforward comparison with other methods 
we have added these in quadrature to give a net uncertainty $\pm 0.2$.  
This distance is based on a calibration of the Cepheid 
PL relation in the LMC, and assumes 
an LMC distance modulus of $(m-M)_0 = 18.50$ \citep{freedman+01}, the
widely agreed-on contemporary value \citep{schaefer08}.

\subsection{TRGB} 

The red giant branch tip method relies on observationally and theoretically
well understood characteristics of Population II stars, 
which are the most abundant stars in an elliptical galaxy. 
The maximum luminosity (tip) of the red giant branch represents 
the core helium ignition stage of low-mass stars, which happens at approximately
the same core mass.  
For ages in the range $\sim 2-15$ Gyr this implies that  
the bolometric luminosity of the tip for stars with the same metallicity 
varies by at most 0.1 mag, resulting in a sharp discontinuity in the 
luminosity function (LF) of the red giant branch at that point. This 
discontinuity in the LF was established in globular 
clusters and dwarf galaxies by \citet{DA90} and \citet{lee+93}, who
provided the first calibration and detection method for the 
TRGB as distance indicator.
In fact, these papers represent the modern re-definition of a
standard candle that was used long ago for globular clusters
in the Milky Way \citep[the mean luminosity of the brightest
giants; see, for example,][]{shapley18} and was a key element
in establishing the size and structure of the Galaxy.

When used in objects containing a significant population of old and 
relatively metal-poor stars, TRGB is a clean and powerful standard candle.   
Observationally it relies only on obtaining single-epoch accurate 
photometry of a large sample of stars in a halo field. Another advantage 
is that reddening or differential reddening for halo stars is not 
usually a significant problem in the analysis, as opposed to Cepheids,
which are typically located in star forming regions with patchy extinction.  
%The TRGB technique is also completely independent of Cepheid, 
%PNLF, SBF and LPV distance determination methods.  
See \cite{rizzi+07} for valuable and more extensive summary of the characteristics
of the method.  

The TRGB method can be used in either optical or near-infrared
wavebands, but is most widely established in the $I$-band.  
In the $I$ band, the metallicity dependence of the 
RGB tip luminosity is small: in particular, for stars with metallicity 
below [Fe/H] = -0.7, the TRGB is essentially ``flat'' in $M_I$
and has been well calibrated both 
theoretically and experimentally through a combination of
stellar evolution models and the data for the nearby extremely
rich Milky Way globular cluster $\omega$ Centauri.
\citet{bel01} have refined the model-based calibration of 
this technique, obtaining
$M_I^{TRGB} = -4.04 \pm 0.12$ for stars in this metal-poor range.
An additional thorough update of the calibration \citep{rizzi+07} which uses 
a combination of RGB data in Local Group dwarf galaxies finds
$M_I^{TRGB} = -4.05 \pm 0.10$ for this metal-poor side of the RGB tip.
In practice, what this means is that {\sl for any galaxy or star cluster
whose stellar population has a strong metal-poor component}, the
RGB tip will be resolved at this luminosity level.
Since the halo stars of NGC 5128 cover the entire metallicity
range from [Fe/H] $\simeq -2.0$ up to above Solar abundance
\citep{rej05}, this condition is easily satisfied.  
Thus the simplest application of the method is to determine the tip
magnitude at the metal-poor side in $I$ where it first resolves,
and subtract the absolute magnitude.
%

% It is preferable to embed your figures in the text as in the following example
\begin{figure}
%\begin{center}
%\includegraphics[scale=0.5, angle=0]{cmd_trgb.eps}
\resizebox{\hsize}{!}{
\includegraphics[angle=0]{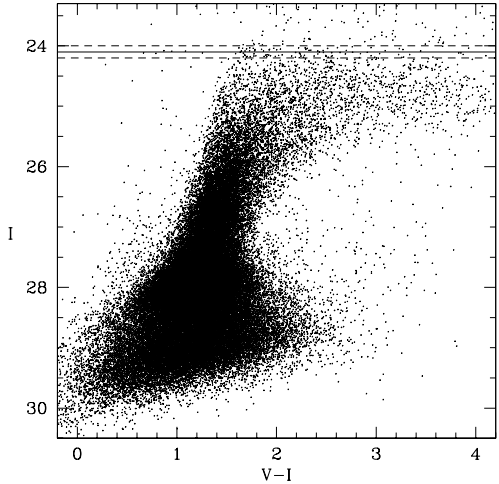}
}
\caption{Color-magnitude diagram for the outer halo of NGC 5128,
from \citet{rej05}.  The photometry is from HST ACS/WFC
imaging in $V$ and $I$ and includes roughly 70000 stars at a
location 40 kpc from the galaxy center.  The horizontal solid
line at $I = 24.1$ shows the deduced luminosity of the TRGB, with
dashed lines showing the measurement uncertainty of $\pm 0.1$ mag. }\label{fig1}
%\end{center}
\end{figure}

$(V,I)$ photometry of resolved stars in three halo fields 
(at projected galactocentric distances of $\sim$20, 30, and 40kpc) 
has been done in NGC 5128 with the HST cameras WFPC2 and ACS/WFC
\citep{har99,har00,har+02,rej05}.  The results for the 40-kpc field, which
is the deepest of the three, are shown in Figure \ref{fig1}.
We have redone the TRGB calculations for all three fields using the 
smoothed probability density function LF methods 
described in those papers, and find $I^{TRGB} = 24.10 \pm 0.10$ for 
each of these outer-halo fields {\sl at the blue, metal-poor side
of the CMD}.
Combined with the most recent Rizzi et al. calibration and using 
$A_I = 0.22 \pm 0.02$ \citep{schlegel98}, with reddening law from
\citet{cardelli89}, we therefore find $(m-M)_0 = 27.93 \pm 0.13$
for the average of the three regions. 

\cite{rizzi+07} provide a thorough discussion 
of the TRGB zero point and application of the method to several nearby galaxies.
As part of their analysis they derive several empirical equations
that describe
the dependence of the zero-point of the TRGB luminosity on the 
metallicity (or color) of the stellar population, in various filters
including $(I, J, H, K)$ and the HST flight system $F814W$.
Their analysis confirms that in $I$ or $F814W$ the metallicity 
dependence of the tip luminosity
is quite modest, and virtually flat for [Fe/H]$ < -1.2$.
They also remeasured the tip magnitude for NGC 5128, finding
$I^{TRGB} = 24.03 \pm 0.02$ (internal uncertainty).  This estimate
nominally agrees well with ours within the uncertainties of both.  However,
they appear to have used the raw WFPC2 data for the inner-halo 8 kpc field
\citep{har+02}, which is by far the most affected by crowding and the
least suitable of the four available halo pointings in NGC 5128.
The much deeper and cleaner ACS data, covering a wider field of view 
(see Fig.~1), give a considerably sharper
definition of the tip magnitude, as do the two other outer-halo fields
that we have used here.  In addition, Rizzi et al. apply
$M_I^{TRGB}=-3.90$, a value more appropriate for the {\sl red, metal-rich}
side of the RGB tip rather than the metal-poor side that represents
the true onset of the RGB tip in any optical bandpass.

\citet{madore+09} discuss in detail the modified 
detection method for measuring the luminosity of the TRGB in 
composite stellar populations (such as galaxy halos) accounting 
explicitly for the metallicity (or color) dependence of the slope
of the tip magnitude. In this new method the impact of reddening is
further reduced. 

While much less used in the near-infrared, TRGB has the advantage of an
even more reduced dependence on reddening. However, the larger 
dependence of the tip magnitude on metallicity in $J$, $H$ and $K$ bands
with respect to $I$ (see also Fig.2 of Rizzi et al.), 
has made this method less popular in the near-infrared.
Using deep $JHKs$ ground-based observations of a halo field at a projected 
distance of 18 kpc north-east from the center, \citet{rej04} 
found $(m-M)_0 = 27.89\pm 0.20$ (J), $27.90\pm 0.20$ (H), 
and $27.88\pm 0.16$ (K). The calibration adopted for the $Ks$ band is based on 
an empirical relation of RGB tip magnitude as a function of metallicity 
for Galactic globular clusters \citep{ferraro+00}, which is in agreement
with the most recent and first {\it geometric} calibration of the TRGB absolute 
magnitude at $M_K = -6.85 \pm 0.03$
based on Hipparcos parallaxes of Solar-neighborhood 
Galactic red giants \citep{tabur+09}.
The $J$ and $H$ band calibrations have somewhat larger uncertainty and
are based primarily on stellar evolutionary models \citep{bertelli+94}
and an estimate of the Galactic Bulge TRGB magnitude \citep{zoccali+03}.
A simple weighted average of these three near-infrared TRGB distance moduli
gives $(m-M)_0 = 27.89\pm 0.11$.

Averaging together the optical and near-infrared results,
we obtain $(m-M)_0 (TRGB) = 27.91 \pm 0.08$.

\subsection{PNLF}

It was shown in the late 1970s that the PNLF is an excellent standard candle 
for distance measurement to nearby galaxies
\citep[see for example,][for a comprehensive recent discussion]{cia02,cia03}.
Logistically similar to the TRGB method, it employs intermediate-age and old
stellar populations and relies on the sharply defined upper end of the PNLF, 
$M^{\star}$, as measured in the [OIII] emission line. Technically, the PNLF 
distance is derived by fitting the observed LF to an empirical
law, and calibrating the observed magnitude with respect to PNLF cutoff magnitude
in M31. The PNLF method is well calibrated within the Local Group and has 
been shown to be consistent in different group and 
galaxy environments.  It is also particularly
helpful in bridging the Population I and II
distance scale techniques, since PNe can be commonly found in
both spiral and elliptical galaxies \citep{cia02}.  

\citet{hui95} determined the PNLF 
for  more than 200 planetary nebulae in NGC 5128 and found 
a distance by matching with the PNLF 
data in M31 and using the then-standard fiducial M31 
distance of 710 kpc ($(m-M)_0 = 24.26$).   
Combining the uncertainties due to the M31 distance, 
the PNLF model, and the filter calibration they quote
a final best-estimate distance modulus for NGC 5128 of 
$(m-M)_0 = 27.73 \pm 0.14$.   
However, the current nominal value 
for the distance to M31 is 775 kpc ($(m-M)_0= 24.45 \pm 0.07$), 
which is based on the Cepheid distance modulus 
\citep{fer00}, the more recent TRGB value \citep{mcc05},
and RR Lyraes \citep{brown04,sara09}, all of which now agree 
to within 0.1 magnitude.
This revision brightens the fiducial PNLF luminosity to
$M^{\star} = -4.67 \pm 0.07$.  
From the numbers in Hui et al. we find that the apparent 
magnitude of the PNLF tip is $m^{\star} = 23.25 \pm 0.07$
and the resulting distance
modulus to NGC 5128 is then $(m-M)_0 = 27.92 \pm 0.12$. 

\subsection{Mira variables (LPV)}

Miras are large amplitude, long-period variables (LPVs), at the tip of the
asymptotic giant branch (AGB). The classic long-period Mira variable stars, 
with periods shorter than $\sim 400$~days, 
follow a well defined infrared and bolometric PL relation
\citep{feast+whitelock99,whitelock+08}. The Mira
PL relation was first found from statistical parallax work, but only with
observations in near-infrared did it become clear that the scatter of this 
relation in the $K$-band (and in bolometric magnitude) is
small enough to be useful for distance determination to other galaxies
\citep[$\sigma_K=0.13$ and $\sigma_{Mbol}=0.16$;][]{feast+whitelock99}. 
The latest calibration of the Mira PL relation is based on the analysis of the 
Hipparcos data for large-amplitude AGB variables in the Milky Way
\citep[for detailed discussion see][]{whitelock+08}. 
Other calibrations that reinforce the definition of the PL relation
use the Milky Way globular clusters
and the LMC Miras.  These additional data therefore rely on RR Lyrae and subdwarf 
main-sequence-fitting distances for globular clusters, and on the variety of the 
sources for the LMC distance moduli. 

This method is strongly reminiscent of the Cepheid method, and like
the Cepheids, the relation is well calibrated from within the Local Group
(as noted above, primarily from the Milky Way and LMC).
Another similarity with Cepheids is the possible (but in the $K$-band small and
debated) metallicity dependence of the PL zero-point \citep{feast04}. 
On the other hand, use of the infrared photometry minimizes errors due to
uncertain extinction (and reddening law).

\citet{rej04} measured several hundred long-period variable stars 
to construct a PL relation for the Miras
in two halo fields of NGC 5128.  The slope of the relation is the same
as for Miras in the LMC and the Milky Way to well within the internal
uncertainties of all.  The zeropoint of the PL relation is
taken from the LMC for which (as above for the Cepheids) $d \equiv 50$ kpc
($(m-M)_0 = 18.50$).  The resulting distance modulus for NGC 5128 based on 
Miras quoted by \citet{rej04} is $(m-M)_0 = 27.96 \pm 0.11$.

\subsection{SBF}

The surface brightness fluctuations method, pioneered by Tonry and 
collaborators in the late 1980s, uses the spatial fluctuation signal in the 
smooth integrated light of the brightest component of the stellar population  
to determine the distance to its parent galaxy \citep[see][and references
therein for discussion of the method]{ton01}. In an old stellar population
typical for elliptical galaxies, measurements in the $I$ band are 
dominated by the red giant branch stars, and while the absolute fluctuation
magnitude is expected to vary with age and metallicity of a stellar population, 
its mean color can be used to constrain the distance measurements to
$\sim 10$\% accuracy \citep{bla09}.
Tonry et al. quoted $(m-M)_0 = 28.12 \pm 0.14$ for NGC 5128.  
The SBF method is best applied to galaxies with smooth bulge light
dominated by an old stellar population (S0 and
ellipticals), but its calibration has so far relied heavily on 
Cepheids.  NGC~5128 is actually the first (and only)
elliptical galaxy where both methods may be applied. So far, the calibration from 
Cepheids used either the association between SBF distances to spiral galaxy bulges
for those spiral galaxies that have Cepheid distances, or SBF distances to
ellipticals and S0 galaxies in groups that host spirals with Cepheid distances
\citep{ton00,bla02}. Efforts to improve the understanding of the SBF magnitudes as
a function of stellar population, and SBF measurements in Magellanic Cloud
clusters, have led to re-calibrations of the SBF zero point.

An inherent uncertainty of the SBF technique is that it  
requires knowledge of the mean colour of the underlying population to account for
metallicity and stellar population differences. In the case of NGC~5128 this
is more difficult to determine because it is so nearby and spread out across
the sky. Additionally, differences in internal  
reddening and an age mixture of populations may be present.
These difficulties are emphasized as well by \citet{fer07}. 
New optical wide-field observations of the galaxy \citep{peng04}, 
combined with the updated zero point of the SBF distance scale, led to the 
revised SBF distance to NGC~5128 of 
$(m-M)_0 = 27.74 \pm 0.14$, reported by \citet{fer07}. 
We adopt here this SBF distance value.

The SBF method requires additional comment because it is the only
one that is difficult to calibrate from {\sl Local Group 
members alone}.  The other four can be thought of as ``primary'' 
techniques in the sense that they use the properties of {\sl directly resolved
individual stars} and ultimately rest on well understood stellar
physics: the Miras and Cepheids both have well defined PL relations (both 
with calibrations made with parallax methods),
while the TRGB and PNLF use the fiducial luminosity at a
breakpoint along the evolutionary path of old stars.
SBF is a ``secondary'' method in the sense
that it relies on the integrated light of an ensemble of stars and must
use the primary methods to set its zeropoint.  
Recently the SBF zero point now has a
calibration using TRGB distance measurements \citep{mould+sakai09}.

Lastly, TRGB and SBF both use the same fundamental stellar population -- the old 
RGB stars -- and thus in principle they should give the same distances for the 
same galaxy.  The main difference between them is, again, that TRGB uses the
resolved stars at the bright tip of the red giant branch, while SBF measures 
the mean luminosity of the entire unresolved or partially resolved RGB population.
NGC 5128 is one of a very small number 
of galaxies beyond the Local Group in which we can make 
a direct comparison between the two.  As can be seen from the summary table, 
although they agree formally (within the quoted uncertainties), 
the resulting distances differ by $~0.3$ Mpc or almost 10\%. 

We can make a similar comparison of TRGB and SBF distances 
for the Leo Group ellipticals 
NGC 3377 and NGC 3379, which at $d \sim 10$ Mpc are close enough that 
HST/ACS imaging can resolve their 
brightest halo stars.  For NGC 3377, \citet{har07a} 
find a TRGB distance modulus of $30.18 \pm 0.16$ which 
is remarkably close to the SBF value of 
$30.19\pm 0.09$ given in Tonry et al. (2001).  
The agreement in NGC 3379 is nearly as good where \citet{har07b} find 
$(m-M)_0 = 30.1 \pm 0.16$ for the TRGB, compared with the Tonry et al. 
SBF value of $30.06 \pm 0.11$.  This suggests to us that the 
much greater difference between the NGC 5128 SBF and TRGB 
distances may be due to factors such as uncertainties 
in the colour of the underlying galaxy light as already mentioned above, or the 
partial presence of a younger stellar component (brighter than the RGB tip), 
which would make it appear brighter and therefore closer \citep{raimondo+05}. 
The latter possibility warrants particular attention since 
the thermally pulsing AGB component (TP-AGB) is surely
present in NGC~5128 as testified by Mira LPVs \citep{rej+03}. 
 
By contrast, the TRGB luminosity is
much less ambiguous since it is based on clearly
resolved stars, has been determined for several 
NGC 5128 halo fields, and is well normalized for 
metallicity.  Thus NGC 5128 is a galaxy 
for which the TRGB distance 
can be argued to supersede the SBF distance.

\section{Discussion and Conclusions}

The extragalactic distance scale is a classic astronomical
subject whose roots extend far into the past.
It will already be evident from the discussion in preceding sections that
the calibrations and zeropoints of standard candles are a constant
source of concern, revision, and debate in the distance scale literature.
With the appearance of new distance calibrators, 
several specialist conferences have been dedicated to this 
topic in the last two decades.
Among the newcomers in the group of well established distance candles
over the last $\sim 20$ years are TRGB, PNLF, and SBF, three of the five distance
indicators used to determine distance to NGC~5128.

%%Format tables as in the following example
\begin{table}
\begin{center}
\caption{Summary of Distance Calibrations}\label{table}
\begin{tabular}{lc}
\hline Method & $(m-M)_0$ \\
\hline
\\
Cepheids         &  $27.67 \pm 0.20$ \\
TRGB (I), (JHK)  &  $27.91 \pm 0.08$ \\
PNLF             &  $27.92 \pm 0.10$ \\
LPV (Miras)      &  $27.96 \pm 0.11$ \\
SBF              &  $27.74 \pm 0.14$ \\
\\
$\langle m-M \rangle_0$ & $27.91 \pm 0.05^a$ \\
\\
$\langle d \rangle$ (Mpc) & $3.82 \pm 0.09^a$ \\
\hline
\end{tabular}
\medskip\\
$^a$Average of the first four methods listed (see text).\\
\end{center}
\end{table}

Although we have stressed the advantages of combining independent approaches,
none of our four key resolved-star standard candles can ultimately
be said to be totally independent of all the others.  The distance scale
for the nearby universe requires the careful assembly and intercomparison
of many techniques, with checks and balances at every outward step.
Convergence may at times seem remarkably slow and difficult.
Nevertheless, at this stage in
the development of the subject, wide agreement to within $\pm 0.1$ mag
has been achieved for galaxies within the Local Group.
We use this ``near-field'' Local Group region as our starting point.

Brief comparisons of the basis of each method show how they
are interrelated at their starting points and how none are
truly ``independent'' of the others.  But
These same points of overlap
provide several strong consistency checks:

\noindent (i) 
The fundamental calibration of the Cepheid PL relation relies 
on trigonometric parallaxes of nearby Cepheids and main-sequence
fitting to young Milky Way star clusters containing Cepheids, often
supplemented by Baade-Wesselink method parallaxes and
the PL slope from the LMC \citep[for only a recent sampling of the vast literature,
see][]{van07,an07,gro08}. 

\noindent (ii) The TRGB luminosity calibration \citep[well reviewed
by][]{rizzi+07} relies on observations of the tip luminosity in
Milky Way globular clusters (particularly $\omega$ Centauri) and
the old-halo components of Local Group dwarf galaxies, all of whose
distances rely in turn on the luminosities of the old RR Lyrae and
horizontal-branch stars.  

\noindent (iii) The calibration of the Mira PL relation
rests on local Milky Way long-period variables and importantly on
the LMC LPVs; the LMC fiducial distance relies in turn on a mixture
of indicators including RR Lyraes, Population I and II Cepheids,
and the expansion-shell parallax to SN1987a.  

\noindent (iv) Finally, the PNLF calibration
that we adopt here depends strongly on the distance to M31, which
in turn is calibrated from Cepheids, Miras, RR Lyraes, and TRGB.

As soon
as we step beyond the Local Group, measurements of distances to
individual galaxies are occasionally affected by larger disagreements
among the methods that are sometimes still hard to understand (see the papers
above for several examples).  For NGC 5128, however, the various methods
have finally begun to converge, showing encouraging agreement to within their
internal uncertainties.  It is worth noting that the true uncertainty
applicable to each method is currently dominated by the {\sl external}
accuracy of the calibration, which is $\pm 0.1$ mag or greater.  
The {\sl internal} measurement uncertainty
-- e.g. the apparent magnitude of the RGB tip, or $M*$ for the PNLF --
is now below 0.1 mag thanks to large observational samples and
rigorous numerical analysis methods.

Table 1 summarizes our results.  The uncertainty quoted for
each one is the combination of internal and external errors.
A simple weighted mean of all five methods
gives $(m-M)_0 = 27.89 \pm 0.04$ or $d=3.77 \pm 0.08$ Mpc. 
Given the discussion above, however, we recommend a final
average based on the four primary, 
resolved-star methods (Cepheids, TRGB, PNLF, Miras).
This average gives a slightly larger distance of $(m-M)_0 =27.91 \pm 0.08$ or 
$d=3.8 \pm 0.1$ Mpc.   

The mutual agreement among these methods is about 
as good as we have for any galaxy beyond the Local Group.  
It appears that, to within $\pm 0.1$ Mpc, 
the recommended distance of 3.8 Mpc for NGC 5128 
is well supported by the evidence at hand.

\section*{Acknowledgments} %If needed
GLHH and WEH acknowledge financial support through research 
grants from the Natural Sciences and Engineering Research Council of Canada.  
The germ of this paper was initiated during a visit to ESO Garching, sponsored
through the ESO visitor programme.

%\end{multicols}

\end{document}